\documentstyle[11pt,newpasp,twoside,epsf]{article}
\def\cm{{\rm\thinspace cm}}

\def\erg{{\rm\thinspace erg}}

\def\s{{\rm\thinspace s}}

\def\ergpcmsqps{\hbox{$\erg\cm^{-2}\s^{-1}\,$}}

\def\ergps{\hbox{$\erg\s^{-1}\,$}}

\def\psqcm{\hbox{$\cm^{-2}\,$}}

\def\edcomment#1{\iffalse\marginpar{\raggedright\sl#1\/}\else\relax\fi}
\reversemarginpar

\begin{document}
\title{Obscured Accretion and Black Holes}
 \author{A.C Fabian}
\affil{Institute of Astronomy, University of Cambridge, Madingley
Road, Cambridge CB3 0HA, UK}

\begin{abstract}
The evidence for obscured AGN and in particular for obscured quasars
is discussed. The spectrum and source modelling of the X-ray
Background suggests that most massive black holes grow by obscured
accretion. A possible major growth phase which is Compton thick is
explored and shown to be difficult to detect directly with current
instruments.

\end{abstract}

\section{Introduction}

Obscured Active Galactic Nuclei (AGN) are required by most models for
the origin of the 1--40~keV X-ray Background (XRB: Setti \& Woltjer
1989; Madau et al 1994; Comastri et al 1995). Here I am interested 
in whether a) most massive black holes underwent an
intrinsically-luminous, but obscured, phase and b), if so, was the
absorption Compton-thick (i.e. column density $N_{\rm H}>1.5\times
10^{24}\psqcm$). In other words, did most massive black holes grow by
obscured accretion?

I rely on the usual observational handles of the XRB, namely its
spectrum, the extragalactic source counts, source identifications and,
as a constraint, the infrared background.

As a hint that highly obscured AGN might be common, Matt et al (2000)
point out that the 3 nearest AGN with X-ray luminosities above
$10^{40}\ergps$ (NGC4395, the Circinus galaxy and Cen A) are all
highly absorbed, with the first two being Compton thick. The main
issues here though are whether there are much more distant examples,
and whether Compton-thick quasars exist.

\section{Clues from the spectrum}

The shape of the spectrum of the XRB is now well known from HEAO-1,
ASCA and BeppoSAX, even if there remains some uncertainty about its
normalization. In the 2--7~keV band it fits a power-law with photon
index 1.4 (e.g. Gendreau et al 1995) and it peaks (in $\nu I_{\nu}$)
at about 30~keV (Marshall et al 1980). The flatness of the spectrum is
only plausibly made by summing many absorbed sources. If these sources
have typical intrinsic quasar spectra with a photon index $\Gamma=2$
then most ($>85$ per cent) of the accretion power must be absorbed
(Fabian \& Iwasawa 1999). Correction of the XRB spectrum for the
minimum absorption necessary reveals the intrinsic mean energy density
of radiation from accretion, which through $L=0.1 \dot M c^2$, yields
the mean local mass density in black holes. The result (Fabian \&
Iwasawa 1999) agrees with observations of that density, inferred from
the motions of stars in galactic nuclei (Merritt
\& Ferrarese 2001).

To obtain the flat spectrum of the XRB all the way to its 30~keV peak
requires that the sources dominating above about 10~keV are Compton
thick. The major uncertainty in the translation from mean energy
density to mean black holes mass is the ratio of the mean redshift to
the radiative efficiency of accretion $(1+z)/\eta$. In the above
$z\sim 2$ and $\eta\sim 0.1$. 

\section{Source identifications}

In the medium hard band of 2--7~keV, the XRB has been resolved by
Chandra (Mushotzky et al 2000; Barger et al 2001; Brandt et al 2001;
Giacconi et al 2001) and XMM (Hasinger et al 2001). Bright source
identifications in the harder 5--10~keV band have been done from
BeppoSAX by the Hellas project (Comastri et al 2000). 

Much has been said at the meeting about the faint sources found in
deep exposures. Since however the 2--7~keV counts turn over around
$10^{-14}\ergpcmsqps,$ source numbers are maximised per unit exposure
time by using short, 10--20~ks exposures with Chandra and XMM. The
sources thereby found are highly relevant to the XRB since more than
half the background originates above that flux.

We (Fabian et al 2000; Crawford et al 2001a; Cowie et al 2001;
Crawford et al 2001b) have been identifying serendipitous hard
sources found in my Chandra cluster fields as a means to study this
population. Some of the clusters, e.g. A2390, have been well-studied
at other wavelengths, and with strong lensing enable us to go much
deeper in all bands. Of particular note is that the A2390 field was
the deepest exposure made with ISO.

The total exposure of the A2390 field is 19~ks and we have made 22
optical identifications (Crawford et al 2001b; Cowie et al 2001). Of
13 sources with spectroscopic redshifts most have soft X-ray spectra.
These are Seyferts and quasars. A further 4 sources have photometric
redshifts. Of the remaining sources 7 are hard and faint. The optical,
X-ray and infrared spectra of two obscured quasars are shown in
Figs.~1 and 2. Both are lensed by the cluster resulting in one being
magnified by a factor of about 2 and the other by 8; both were
detected by ISO (Lemonon et al 1999). The less magnified one (source
A18) has an intrinsic 2--10~keV luminosity of about $3\times
10^{45}\ergps$ and $N_{\rm H}\sim 3\times 10^{23}\psqcm$ (Fig. 1).

\begin{figure}
\plottwo{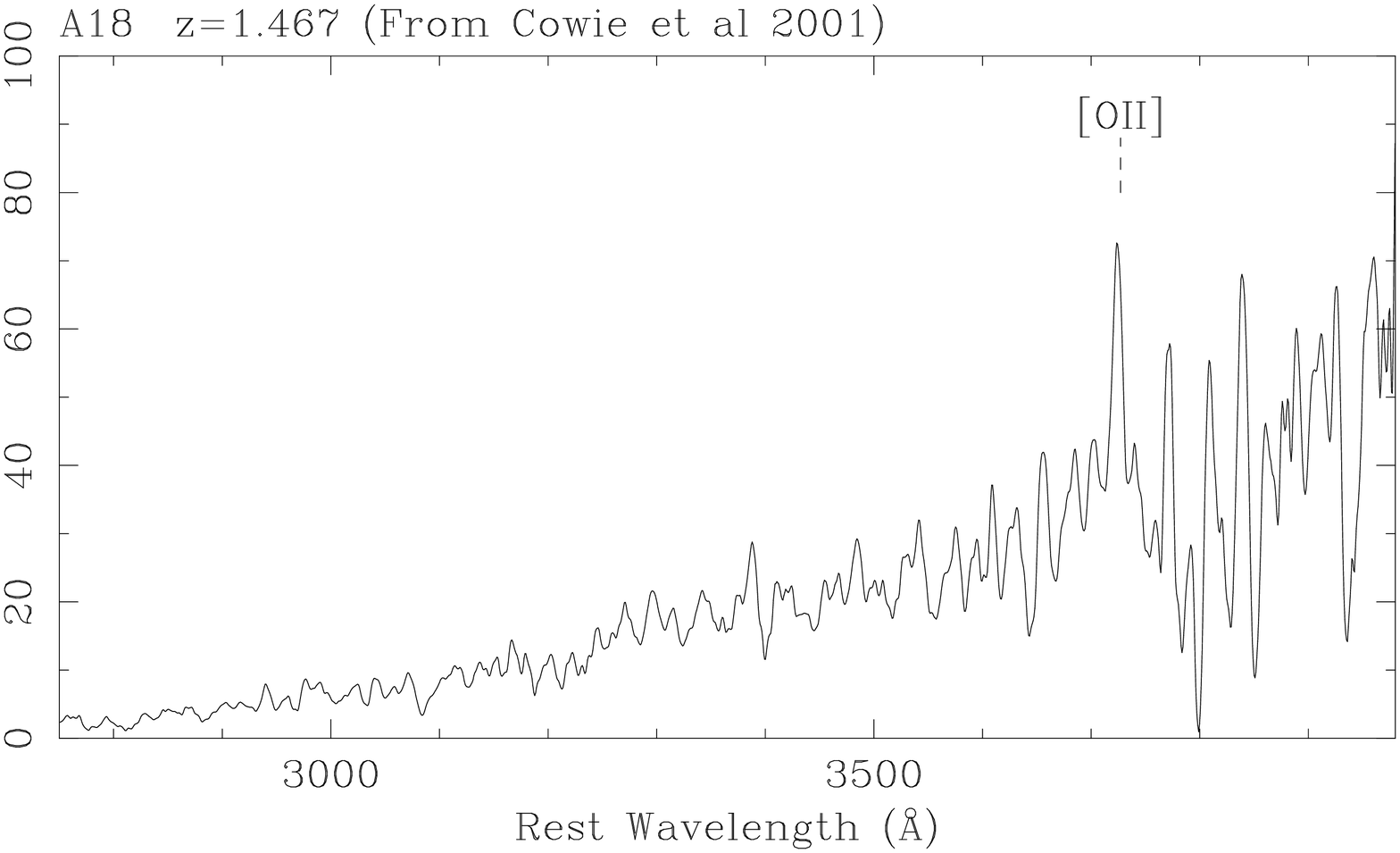}{a18nir.epsi}

\plottwo{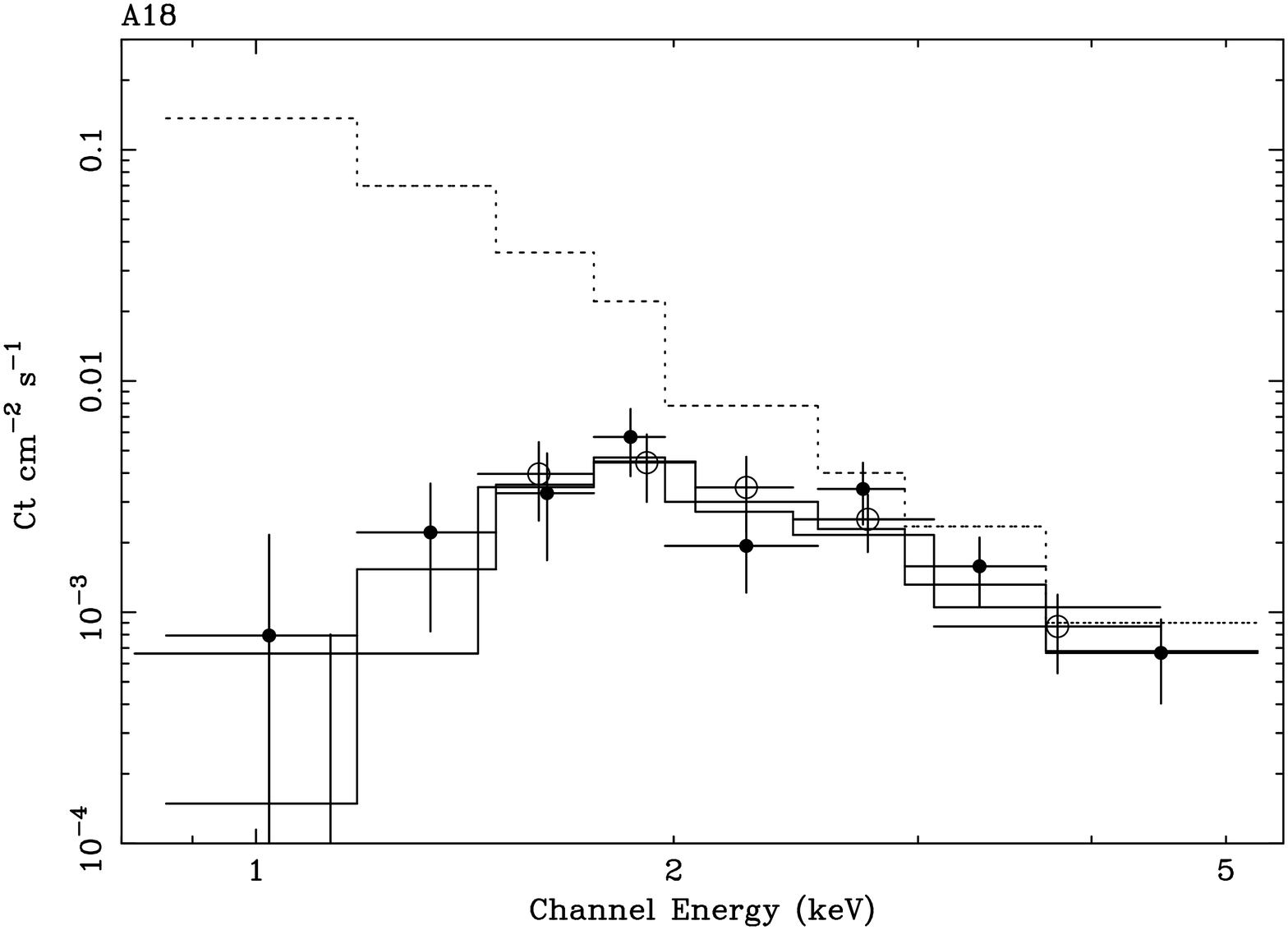}{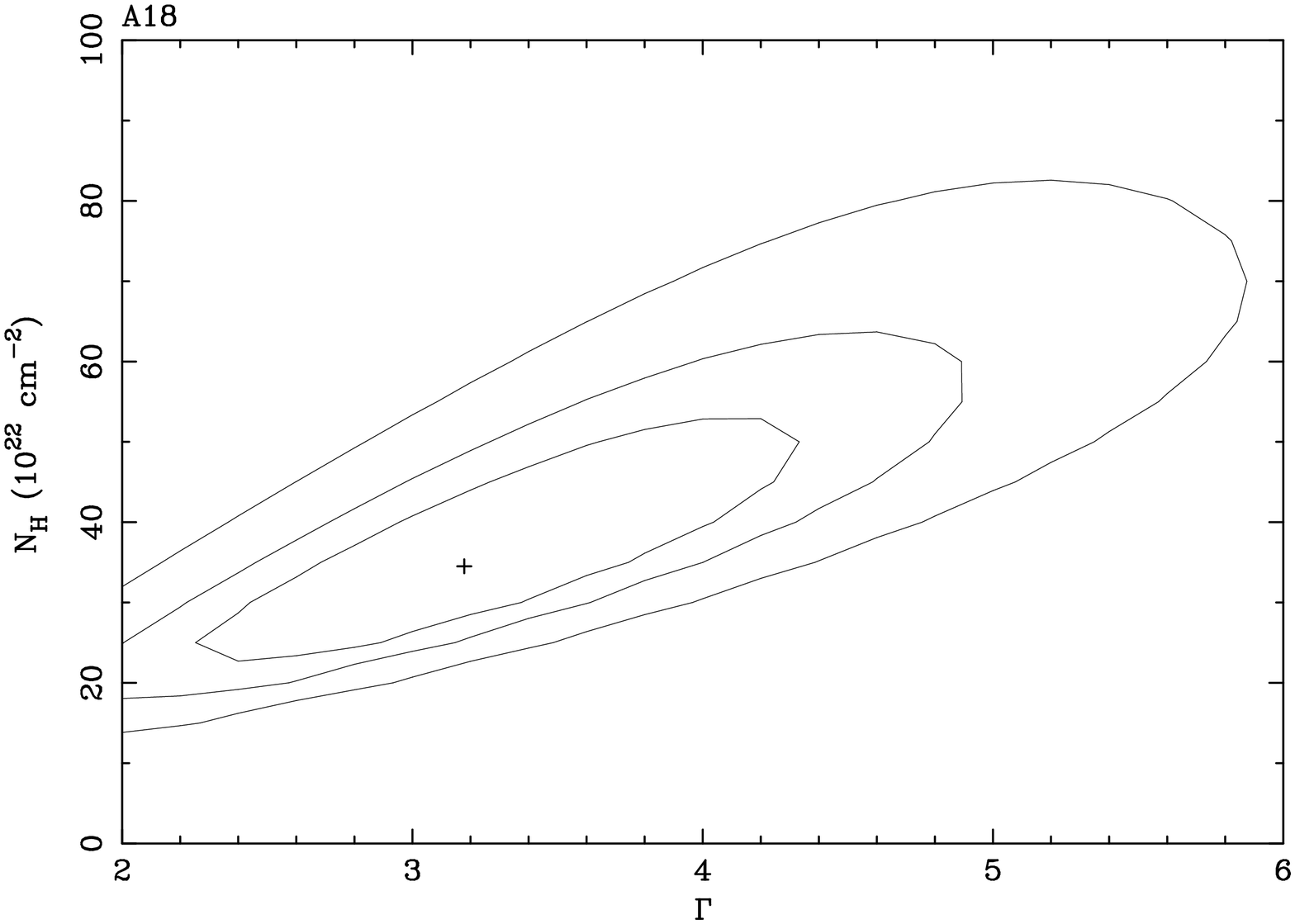}
\caption{Spectral results for source A18 in the A\,2390 field.
Clockwise from top left is shown the optical spectrum with [OII]
emission (Cowie et al 2001), optical and near IR photometry with a
HYPERZ galaxy fit, Contours of photon index $\Gamma$ and intrinsic
column density (the source is at $z=1.467$), and finally the X-ray
spectrum (the dashed line indicates what the spectrum would look like
if the source had no intrinsic absorption). The last 3 panels are from
Crawford et al (2001). The source is a powerful Type II quasar, with
an intrinsic 2--10~keV luminosity of $3\times 10^{45}\ergps$.}
\end{figure}

The 15\,$\mu$m ISO flux shows that the absorbed X-rays and UV emission
are reradiated in the mid-far infrared by warm dust (Fig.~2). The
correlation of serendipitous X-ray sources with SIRTF images should be
very fruitful.

Such work, together with the deep fields and other serendipitous
sources studied by others, shows that obscured quasars exist. We have
of course known for long that many powerful radio galaxies are
obscured along our line of sight, and also, more recently, radio
quasars (the red quasars; Francis et al 2000: BAL radio quasars;
Najita et al 2000, Gregg et al 2001) and more generally BAL QSO
(Gallagher et al 2001). In the unified model they appear as quasars
when looked at closer to the jet direction.One example of a powerful
distant radio galaxy is B2 0902 (Carilli et al 1994) at $z=3.4$.
Chandra finds this to be highly absorbed, by a column of at least
$10^{23}\psqcm$ and an intrinsic 2--7~keV power exceeding
$10^{46}\ergps$ (Fig.~3; Fabian et al, in preparation). Other highly
absorbed objects are IRAS09104 (Franceschini et al 2000; Iwasawa et al
2001) and IRAS F15307 (Fabian et al 1996). Intrinsic absorption is
also seen in several very distant blazars, at $z>4$ (Boller et al
2000; Yuan et al 2000; Fabian et al 2001a,b).

\begin{figure}
\plotone{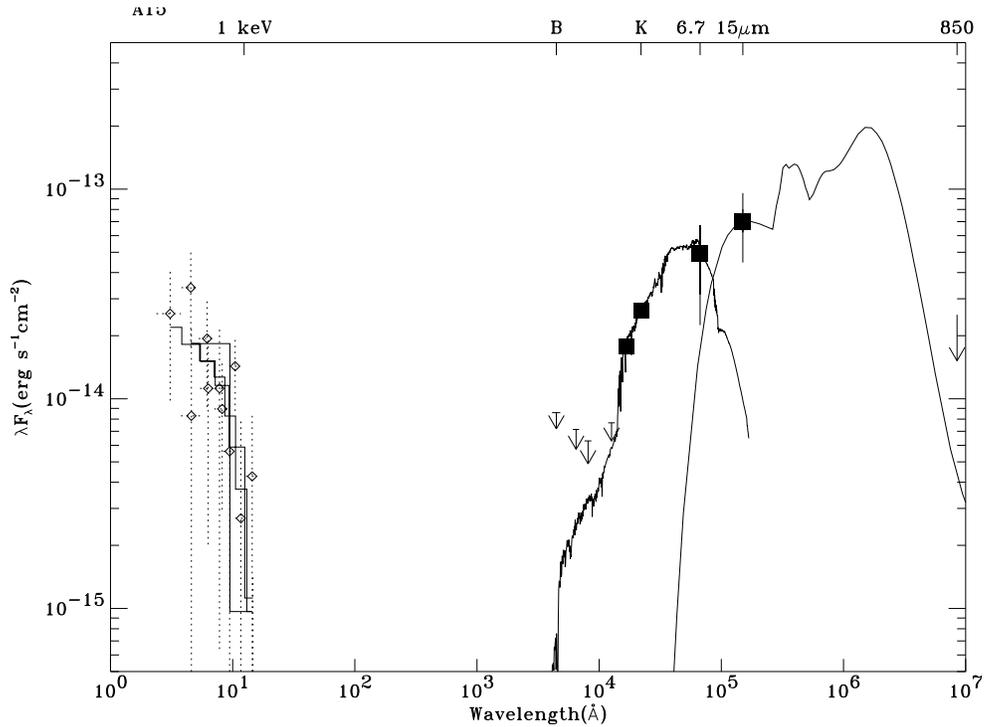}
%\plotfiddle{sed_a15.epsi}{5cm}{90}{0}{0}{0}{0}
\caption{The spectral energy distribution of source A15 in the A\,2390
field. The lensing magnification is about 8. Note the absorbed X-ray
spectrum, the galaxy component (from a HYPERZ fit) between
$10^4-10^5\mu$m and the warm dust component at longer wavelengths.
The source has a photometric redshift of about 2.8 (Cowie et al 2001;
Crawford et al 2001).  }
\end{figure}

\begin{figure}
%\plotone{0902_spectrum.epsi}
\plotone{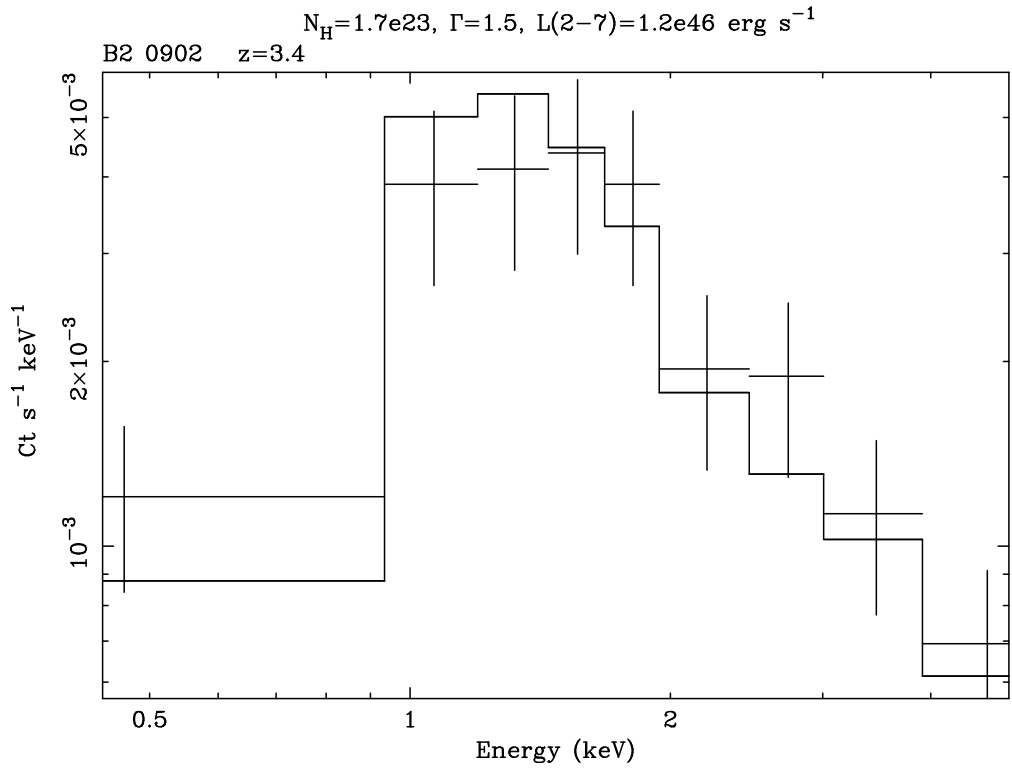}
\caption{The Chandra X-ray spectrum of the $z=3.4$ radio galaxy
B2~0902 (from Fabian et al, in prep). Note the high power and large
absorption.. }
\end{figure}

\section{XRB modelling}

Models for synthesising the XRB can either work backwards in time
using observed (often local) luminosity functions, or forwards in time
using some evolutionary model for accreting black holes and their
galaxy hosts. Examples of the first approach were presented by others
at the meeting. Here I concentrate on the second approach, which may
be preferable for the Compton-thick sources for which no luminosity
function yet exists. 

Wilman, Fabian \& Nulsen (2000) have incorporated a simple obscured
black hole growth model (Fabian 1999) into a semi-analytic model for
galaxy formation (Nulsen \& Fabian 1999). Basically the (isothermal)
galaxy has a significant fraction of its mass in cold, obscuring gas.
A central black hole grows by accretion until its wind has enough
thrust to eject that gas. Force balance leads to the black hole mass
at this stage being proportional to the velocity dispersion of the
galaxy to the fourth power (an earlier model by Silk \& Rees 1998 uses
an energy argument to obtain the fifth power). The gas in the galaxy
is just Thomson thick when ejection takes place. Afterwards the quasar
is unobscured and lasts as long as there is any mass in its disc. The
major growth phase for massive black holes is then before the normal
unobscured phase and thus before $z\sim 1.5$. Some action still
continues after that time from late forming haloes and also if the
black holes are revived by mergers.

The model results are in reasonable agreement with the observed XRB
spectrum and source counts. This does not of course prove that it is
the right answer, and is more a demonstration that such forward
modelling is possible and can yield agreement.

\section{Distant Compton-thick sources}

The above model predicts that much of the peak of the XRB between
10--30~keV is due to Compton-thick quasars. It also predicts the
numbers of such sources which are detectable. The numbers for Chandra
and XMM are rather disappointing, being only a handful per deep
exposure (Fabian et al 2001) and most are within a factor of 1.5 of
the detection threshold. Despite the negative K-correction for
absorbed X-ray sources (Wilman \& Fabian 2000), few sources distant
enough to redshift their emission to a few keV are bright enough to be
detectable.

The net result is that Compton-thick sources are going to be difficult
to find directly, even with powerful instruments such as Chandra and
XMM. They may appear indirectly by scattered emission (e.g. Norman et
al 2001) but their true nature will be difficult to discern.

Resolving the brighter members of the distant, Compton-thick class may
be best done with harder X-ray instruments such as on Astro-E-2,
Constellation-X and Xeus, as well as surveys by LOBSTER and especially
EXIST (Grindlay et al 1999). When this happens, there will again,
justifiably, be the statement that the `XRB has at last been
resolved'.

Much of the power in the XRB is at 10--40~keV. It is difficult to make
the 30~keV peak itself with synthesis models using very distant
obscured AGN, due to the effect of redshift. Enhanced metallicity in
the absorbing gas can somewhat stave off the effects of Compton
down-scattering in the absorber (Willman \& Fabian 2000), but probably
the high energy of the peak is due to a significant contribution by
the lowest redshift Compton-thick Seyferts. If so, the 5--20~keV band
probes the highest redshifts. And the 20--50~keV sky will be rich in
new lower-redshift sources.

\section{Acknowledgements}

I thank Richard Wilman, Carolin Crawford and  Poshak Gandhi for
discussions, Robert Schmidt for technical help, and the Royal Society
for support.

\end{document}